\documentclass[apjl]{emulateapj}

\usepackage[utf8]{inputenc}
\usepackage{graphicx}
\usepackage{amsmath}
\usepackage{natbib}
\usepackage{multirow}
\usepackage{xfp}
\usepackage{xcolor}
\usepackage{ulem}

\newcommand{\agal}{\ensuremath{a_{\rm Gal}}}
\newcommand{\aobs}{\ensuremath{a_{\rm LOS}^{\rm Obs}}}
\newcommand{\amod}{\ensuremath{a_{\rm LOS}^{\rm Mod}}}
\newcommand{\PBDOTobs}{\ensuremath{\dot{P}_b^{\rm Obs}}}
\newcommand{\PBDOTshk}{\ensuremath{\dot{P}_b^{\rm Shk}}}
\newcommand{\PBDOTgal}{\ensuremath{\dot{P}_b^{\rm Gal}}}
\newcommand{\PBDOTGR}{\ensuremath{\dot{P}_b^{\rm GR}}}

\newcommand{\rowify}[2]{\multirow{\fpeval{#1+0.35}}{*}{#2}}

\begin{document}

\title{A measurement of the Galactic plane mass density from binary pulsar accelerations}

\author{Sukanya Chakrabarti\altaffilmark{1,2}, Philip Chang\altaffilmark{3}, 
Michael T.\,Lam\altaffilmark{2,4}, 
Sarah J.\,Vigeland\altaffilmark{3}, 
Alice C. Quillen\altaffilmark{5} }

\altaffiltext{1}{Institute for Advanced Study, 1 Einstein Drive
Princeton, New Jersey
08540 USA; chakrabarti@ias.edu}
\altaffiltext{2}
{School of Physics and Astronomy, Rochester Institute of Technology, 84 Lomb Memorial Drive, Rochester, NY 14623}
\altaffiltext{3}
{Department of Physics, University of Wisconsin-Milwaukee, 3135 North Maryland Avenue, Milwaukee, WI 53211}
\altaffiltext{4}
{Laboratory for Multiwavelength Astronomy, Rochester Institute of Technology, 84 Lomb Memorial Drive, Rochester, NY 14623}
\altaffiltext{5}
{Department of Physics and Astronomy, University of Rochester}

\begin{abstract}
We use compiled high-precision pulsar timing measurements 
to directly measure the Galactic acceleration of binary pulsars relative to the Solar System barycenter.  Given the vertical accelerations, we use the Poisson equation to derive the Oort limit, i.e., the total volume mass density in the Galactic mid-plane.  Our best-fitting model gives an Oort limit of $0.08^{0.05}_{-0.02} M_{\odot}/\rm pc^{3}$, which is close to estimates from recent  Jeans analyses.  Given the accounting of the baryon budget from McKee et al. (2015), we obtain a local dark matter density of $-0.004^{0.05}_{-0.02}~M_{\odot}/\rm pc^{3}$, which is slightly below other modern estimates but consistent within the current uncertainties of our method.  While this first measurement of the Oort limit (and other Galactic parameters) has error bars that are currently several times larger than kinematical estimates, they should improve in the future.  We also constrain the oblateness of the potential, finding it consistent with that expected from the disk and inconsistent with a potential dominated by a spherical halo, as is appropriate for our sample which is within a $\sim$ kpc of the Sun.  We find that the slope of the rotation curve is not constrained by current measurements of binary pulsar accelerations.  We give a fitting function for the vertical acceleration $a_{z}$: $a_{z} = -\alpha_{1}z$; $\log_{10} (\alpha_{1}/{\rm Gyr}^{-2})=3.69^{0.19}_{-0.12}$. By analyzing interacting simulations of the Milky Way, we find that large asymmetric variations in $da_{z}/dz$ as a function of vertical height may be a signature of sub-structure. We end by discussing the power of combining constraints from pulsar timing and high-precision radial velocity (RV) measurements towards lines-of-sight near pulsars, to test theories of gravity and constrain dark matter sub-structure. 
\end{abstract}

\section{Introduction}

By serving as precise astrophysical clocks, pulsars have been used in many tests of fundamental physics \citep[see e.g.,][]{Will2014}. Among these tests, pulsars can enable the detection of the cosmological gravitational wave background \citep[see e.g.,][]{Burke-Spolaor+2019} and provide constraints on the nature of gravity \citep[e.g.,][]{Weisberg2016,Zhuetal2019}.  Here, we explore the idea that pulsars with precisely measured binary orbital periods can serve as effective accelerometers that can be used to directly measure the Galactic acceleration. 

It has been proposed that high precision radial velocity (RV) measurements can be used to directly measure the Galactic acceleration \citep{Silverwood2019, Ravi2019,Chakrabarti2020}.  By quantifying the contamination from planets and binaries to the Galactic RV signal in \cite{Chakrabarti2020}, we showed that even for modest sample sizes, we can reliably expect to extract the Galactic signal by measuring the $\Delta RV$ over a ten-year baseline, despite the presence of planets and binaries in a realistic Galactic population.  Time-dependent potentials as in interacting simulations of the Milky Way lead to asymmetries in the vertical acceleration relative to static models, especially at heights $|z| >~\rm 1 $ kpc relative to the Galactic mid-plane \citep{Chakrabarti2020}.  Prior work has focused mainly on kinematical analysis \citep{Kuijken_Gilmore1989, HolmbergFlynn, Bovy_Tremaine2012, Schutz2018} of various stellar tracers to estimate the Galactic acceleration rather than directly measuring it.  The analysis of an interacting simulation of the Milky Way by \cite{Haines2019} indicates that there are differences in the true density in the simulation relative to that determined from kinematics (such as in the Jeans approximation, which assumes spherical symmetry and equilibrium), especially for perturbed regions of the disk.  In view of the dynamically evolving picture of the Galaxy as manifested by \textit{Gaia} data \citep{helmi2018}, kinematic  estimates should be tested against direct measurements of the acceleration.

Here, we analyze line-of-sight accelerations of fourteen pulsar systems in binaries that have precise measurements of their orbital periods ($P_{b}$) and rate of change in the orbital period ($\dot{P_{b}}$). We determine the radial and vertical Galactic accelerations of binary pulsars, and fit these values as a low-order polynomial as a function of $R$ and $z$ to measure the local potential and its derivatives.  Given these accelerations, we use the Poisson equation to determine the mid-plane density, and accounting for the baryon density from recent work \citep{McKee2015,Bien2014}, we then determine the local dark matter density.  Measurements of the local dark matter density can be used to interpret direct detection measurements of dark matter to ultimately understand the nature of the dark matter particle \citep{Read2014}.  

Pulsar timing has previously been used to infer the potential in globular clusters \citep{Prageretal2017}, and very recently for the Galaxy \citep{Phillips2020}.  The work by \cite{Phillips2020} is contemporaneous with ours.  A key difference in our work arises from our analysis of orbital periods (rather than spin periods), as well as our inclusion of both the vertical and radial components of the acceleration. \cite{Phillips2020}'s value of the acceleration corresponds to a velocity for the local standard of rest $V_{\rm LSR} \sim 350~\rm km/s$.  This value is at odds with the value determined by \cite{Quillen2020} using the Galactocentric radius of the Sun measured by the GRAVITY collaboration et al. (2018), the proper motion of the radio source associated with Sgr $A^{\star}$, and the tangential component of the solar peculiar motion by \cite{Schonrich2010}, which gives $233.3 \pm 1.4 ~ \rm km/s$.  The value in \cite{Quillen2020} is consistent with the measurement using trigonometric parallaxes of high-mass star formation regions from \cite{Reid2019}.  The discrepancy may be due to their statistical analysis of spin periods rather than the direct analysis that can be done for orbital periods.  Measurements of the Galactic acceleration by use of observed spin periods are statistical in nature since they require knowledge of the intrinsic distribution of spin periods and spindowns whereas the use of binary orbital periods do not.  

The current distribution of pulsars with precisely measured $\dot{P}_{b}$ corresponds to approximately a square kpc in area.  A small area coverage like this provides significantly more leverage in measuring gradients in vertical accelerations than radial accelerations.\footnote{After this work was submitted for publication, a similar work appeared by \cite{Bovy2020}.  Contrary to the statement in that work that we determined only “the relative Galactic acceleration at the binary pulsar location and the Sun,” we in fact determine absolute accelerations.  The difference between this work and \cite{Bovy2020} is that we base our results on earlier measurements of $V_{LSR}$, from \citep{Schonrich2012}. In addition, our value of the radial gradient of the rotation curve and \cite{Bovy2020}'s are consistent within the uncertainties.}  Thus, while we solve for both components of the acceleration simultaneously, we will focus here on vertical accelerations.

This paper is organized as follows.  In \S \ref{sec:data}, we review the properties of the pulsars we have selected here, and our method for determining Galactic accelerations from pulsar timing data.  We compare the line-of-sight accelerations of the pulsars to various static models and give the best-fit values in \S\ref{sec:staticMW}.  Here, we also we present our values for the Oort limit, the local dark matter density, and a parameter that is sensitive to the oblateness of the potential.  In \S\ref{sec:discuss}, we compare the results to interacting simulations, and discuss some additional implications of our work.  We summarize our findings in \S\ref{sec:conclusion}.

\section{Analysis and Results}

\subsection{Pulsar Timing Measurements}
\label{sec:data}

We select binary pulsars from the ATNF pulsar catalogue \citep{PSRCAT} that have precisely measured $\dot{P}_{b}$ (non-zero within 2-sigma), distances, and proper motions (either from pulsar timing or very-long-baseline interferometry, VLBI). We do not include pulsars (i) in globular clusters where the additional accelerations induce a change to the observed $\dot{P}_b$, (ii) in systems undergoing ablation or mass transfer that changes the orbital parameters, or (iii) without parameter uncertainties. Our sources along with their parameters are provided in Table \ref{tab:pulsars}; the measurements are given here relative to the solar system barycenter.

For some sources, there are multiple measurements of the observed binary period $\PBDOTobs$ reported. In that case, we choose the data set with lowest uncertainty on $\PBDOTobs$, and use the other timing model parameters from that data set required for our analysis.  Additionally, for some sources, there are multiple measurements of the parallax, e.g., timing parallax and VLBI measurements, and we adopt the parallax value with the lowest uncertainty.  In the case of PSRs J0737$-$3039A/B and J2222$-$0137, where insufficient astrometric information was measured, we used the parallaxes and proper motions derived from VLBI for the purpose of improving gravitational tests with these systems \citep{Deller+2009, Deller+2013}.
Since all of our sources are within $\sim$ kpc of the Sun, we cannot yet probe the global halo potential.  The Hulse-Taylor system \citep{Weisberg2016} is at present the only source that is at a larger radial distance.  We do not include it currently in our analysis as a single source does not help in constraining global potentials, and therefore we focus on the simple potentials we outline below.

\cite{Lorimer_Kramer2004} have discussed the procedure of obtaining astrometric measurements from the times of arrival (TOA) of the pulses, and these measurements have been compared to VLBI measurements \citep{Chatterjeeetal2009,Deller+2019}, and found to be in good agreement.  While there can be potential systematic uncertainties in the TOA analysis due to red noise \citep{Deller+2019}, millisecond pulsars afford the highest precision due to their short frequent bursts and stable rotation.  For pulsars with white-dwarf companions, pulsar timing measurements of the proper motion and parallax have also been compared to \textit{Gaia} parallaxes, and found to agree with \textit{Gaia} parallaxes in general \citep{Jenningsetal2018}.  For sources approaching \textit{Gaia's} limiting magnitude pulsar timing measurements can be more precise than \textit{Gaia} parallaxes \citep{Jenningsetal2018}.  The overall agreement in astrometric quantities derived pulsar timing and other methods (VLBI, \textit{Gaia}) indicates that pulsar timing astrometric measurements are reliable. 


\begin{deluxetable*}{lcccccccl}
\tablecolumns{8}
\tablecaption{Observed Pulsar Parameters
\label{tab:pulsars}}
\tablehead{
\colhead{PSR} & \colhead{$l$} & \colhead{$b$} & \colhead{$\varpi$} & \colhead{$\mu$} & \colhead{$P_b$} & \colhead{$\PBDOTobs$}  & \colhead{$\PBDOTGR$} & \colhead{Reference}\\
\colhead{} & \colhead{(deg)} & \colhead{(deg)} & \colhead{(mas)} & \colhead{(mas/yr)} & \colhead{(d)} & \colhead{($10^{-12}~\mathrm{s}~\mathrm{s}^{-1}$)} & \colhead{($10^{-12}~\mathrm{s}~\mathrm{s}^{-1}$)} &
}
\startdata
J0437$-$4715 & 253.39 & -41.96 & 6.37(9) & 140.911(2) & 5.7410459(4) & 3.728(6) & -0.00273(5) & \citet{Reardon+2016} \\
J0613$-$0200 & 210.41 & -4.1 & 1.25(13) & 10.514(17) & 1.198512575184(13) & 0.048(11) & - & \citet{Desvignes+2016} \\ 
J0737$-$3039A/B & 245.24 & -4.5 & 0.87(14)\tablenotemark{b} & 4.37(55)\tablenotemark{b} & 0.10225156248(5) & -1.252(17) & -1.24787(13) & \citet{Kramer+2006}\\
J0751+1807 & 202.73 & 21.09 & 0.82(17) & 13.7(3) & 0.263144270792(7) & -0.0350(25) & -0.0434(38) & \citet{Desvignes+2016} \\
J1012+5307 & 160.35 & 50.86 & 0.71(17) & 25.615(11)
 & 0.604672722901(13) & 0.061(4) & -0.0109(17) & \citet{Desvignes+2016} \\
J1022+1001 & 231.79 & 51.10 & 1.39(4)\tablenotemark{c} & 17.09(3) & 7.8051360(16) & 0.55(23) & -0.0014(13) & \citet{Reardon+2016}\\
B1534+12\tablenotemark{a} & 19.85 & 48.34 & 0.86(18) & 25.33(1) & 0.420737298879(2)
 & -0.1366(3) & -0.19245(3) & \citet{Fonseca+2014}\\
J1603$-$7202 & 316.63 & -14.50 & 1.1(8) & 7.73(5) & 6.3086296991(5) & 0.31(15) & - & \citet{Reardon+2016}\\
J1614$-$2230 & 352.64 & 20.19 & 1.54(10) & 32.4(5) & 8.68661942256(5) & 1.57(13) & - & \citet{Alam2020} \\ 
J1713+0747 & 28.75 & 25.22 & 0.87(4) & 6.286(4) & 67.8251299228(5) & 0.34(15) & - & \citet{Zhuetal2019} \\ 
J1738+0333 & 27.72 & 17.74 & 0.68(5) & 8.675(8) & 0.3547907398724(13) & -0.0170(31) & -0.0277(17) &  \citet{Freire+2012}\\
J1909$-$3744 & 359.73 & -19.60 & 0.861(13) & 37.025(5) & 1.533449474305(5) & 0.51087(13) & -0.00279(3) & \citet{Liu+2020}\\
J2129$-$5721 & 338.01 & -43.57 & 1.9(9) & 13.32(4) & 6.6254930923(13) & 0.79(36) & - & \citet{Reardon+2016}\\
J2222$-$0137 & 62.02 & -46.08 & 3.742(15)\tablenotemark{d} & 45.09(2)\tablenotemark{d} & 2.44576469(13) & 0.20(9) & -0.0077(4) & \citet{Cognard+2017}
\enddata
\tablenotetext{a}{PSR J1537+1155}
\tablenotetext{b}{Astrometric parameters from \citet{Deller+2009}}
\tablenotetext{c}{Parallax measurement from \citet{Deller+2019}}
\tablenotetext{d}{Astrometric parameters from \citet{Deller+2013}}
\tablecomments{Blank $\PBDOTGR$ entries are either too small or the masses are not known.}
\end{deluxetable*}


For a binary system in the Galaxy not undergoing mass transfer, we may write the observed orbital period drift rate $\dot{P}_{b}^{\rm Obs}$ as: 
\begin{equation}
   \PBDOTobs = \PBDOTgal +  \PBDOTshk + \PBDOTGR,
\end{equation}
where $\PBDOTgal = P_{b} \agal/c$ is the rate induced by the Galactic potential, $\agal$ is the relative line-of-sight Galactic acceleration between the Solar system barycenter and the pulsar,
 $c$ is the speed of light, and $\PBDOTshk$ is the apparent  drift rate caused by the binary's transverse motion (known as the Shklovskii effect; \citealt{Shklovskii}; \citealt{Damour_Taylor91}), which is given by: 
\begin{equation}
\PBDOTshk= 
\mu^2 d \frac{P_{b}}{c}, \end{equation}
for a system at distance $d$ with a proper motion $\mu$. The term $\PBDOTGR$ describes the rate at which the system is losing energy due to gravitational radiation \citep{Weisberg2016}, and can be computed given the orbital period, eccentricity $e$, and the masses of the pulsar $m_p$ and its companion $m_c$ (determined from Shapiro delay; \citealt{shapiro1964}) as
\begin{eqnarray}
    \PBDOTGR &=& -\frac{192\pi G^{5/3}}{5c^5} \left(\frac{P_b}{2\pi}\right)^{-5/3} (1-e^2)^{-7/2} \nonumber \\
        && \times \left( 1 + \frac{73}{24}e^2 + \frac{37}{96} e^4 \right) \frac{m_pm_c}{(m_p+m_c)^{1/3}} \,.
\end{eqnarray}
Given these terms, we can then calculate the line-of-sight Galactic acceleration, $\agal$ as: 
\begin{equation}
\agal = c\frac{\dot{P}_{b}^{\rm Gal}}{P_{b}}.
\end{equation}

We define the observed line-of-sight acceleration, \aobs, as
\begin{equation}
  \aobs= \frac{c \dot{P}_{b}^{\rm Obs}}{P_{b}}.
\end{equation}
This is simply a redefinition of the observed binary period drift rate  $\dot{P}_b^{\rm Obs}$.  As a result, it cannot be compared to a true acceleration as it includes both the Shklovskii effect and secular GR effects, $\dot{P}_b^{\rm GR}$.  Likewise, we also compute a model line-of-sight acceleration, \amod, that includes these additional effects, which we compare to the observed values.  VLBI measurements of the Solar system barycenter \citep{Titov2013} give a value for the solar system acceleration of  $(9.3, 0.4, 0.3) \pm (1.1, 1.1, 1.3)~\rm mm/s/yr$ in the Galactic reference frame, i.e., the vertical component is not statistically significant.  The acceleration of the Solar system barycenter for the models that we consider here are consistent with the VLBI measurements within the uncertainties. 




\subsection{Comparison of pulsar timing data with static models of the Milky Way}
\label{sec:staticMW}

\begin{table}[h]
\begin{center}
\caption{Models,  best-fit  parameters, AIC and reduced $\chi^2$ values}
\begin{tabular}{|l|l|r|r|}
\hline
 Model & Best-fit values & AIC & $\chi^{2}_{\nu}$ \\
 \hline
 $\alpha_{1}$ & $\log_{10}(\alpha_{1}/{\rm Gyr}^{-2}) =3.61^{+0.13}_{-0.10}
 $ & 21 &1.5 \\  
 \hline
 \rowify{2}{$\alpha_{1}, \beta$} &  $\log_{10}(\alpha_{1}/{\rm Gyr}^{-2})=3.69^{+0.19}_{-0.12}$,  & \rowify{2}{22} & \rowify{2}{1.5}\\
 & $\beta = -0.18^{+0.22}_{-0.30}$ & & \\
\hline
 \rowify{2}{$\alpha_{1}, \gamma$} & $\log_{10} (\alpha_{1}/{\rm Gyr}^{-2})=3.77^{0.17}_{-0.10}$, & \rowify{2}{24} & \rowify{2}{1.7} \\ 
 & $ \log_{10}( \gamma/{\rm Gyr}^{-2})=-4.87^{+0.09}_{-0.11}$ & &  \\ 
 \hline 
 \rowify{2}{$\alpha_1, \alpha_2$} & $\log_{10}(\alpha_{1}/{\rm Gyr}^{-2}) = 3.65^{+0.14}_{-0.11} $, & \rowify{2}{29} & \rowify{2}{2.0} \\
&  $\alpha_2 = -279^{+940}_{-215}~\rm Gyr^{-2} kpc^{-1} $ & & \\
 \hline
 \rowify{3}{Local} &  $(da/dr)/(V_{\rm LSR}^2/R_\odot) = -1.3^{+0.45}_{-0.61},$ & \rowify{3}{25} & \rowify{3}{1.7} \\ 
& $(da/d\phi)/(V_{\rm LSR}^2/R_\odot) = -0.16^{+0.59}_{-0.72}$, & & \\ 
& $\log_{10}(da/dz/{\rm Gyr^{-2}}) = 3.73^{+0.20}_{-0.12}$ & & \\
 \hline
 \rowify{2}{$\rho_0 \exp \left(-\frac{|z|}{z_{0}}\right)$} & $\log_{10} (\rho_{0}/M_{\odot}{\rm pc}^{-3}) = -1^{+0.2}_{-0.4}$, & \rowify{2}{43} & \rowify{2}{3.3} \\ 
 & $\log_{10} ( z_{0}/{\rm pc} ) = 3^{+0.9}_{-0.7}$ &  & \\ 
 \hline
 \rowify{2}{Hernquist} &  $M_{h}=0.7^{+1.5}_{-0.5}\times 10^{12} M_{\odot}$, & \rowify{2}{27} & \rowify{2}{1.9} \\ 
 & $a_H = 220^{+1540}_{-183}$ kpc & &  \\ \hline
 MWP & \cite{Bovy2015} values & 25 & 1.8 \\
 \hline
\end{tabular}
\label{tab:Table1}
\end{center}
\end{table}

\begin{figure}[ht]        
\begin{center}
\includegraphics[scale=0.5]{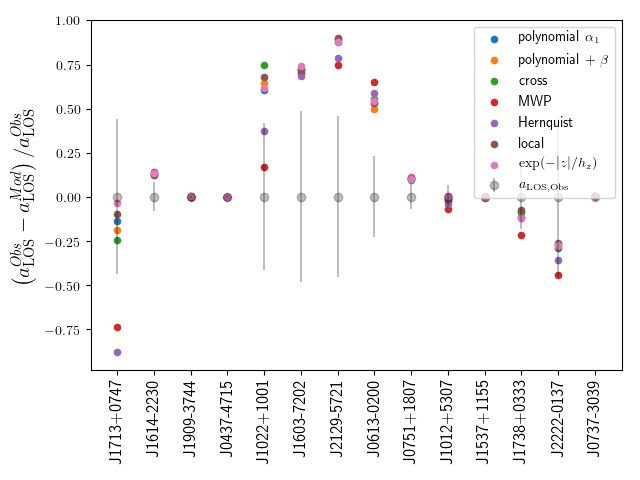}
\includegraphics[scale=0.5]{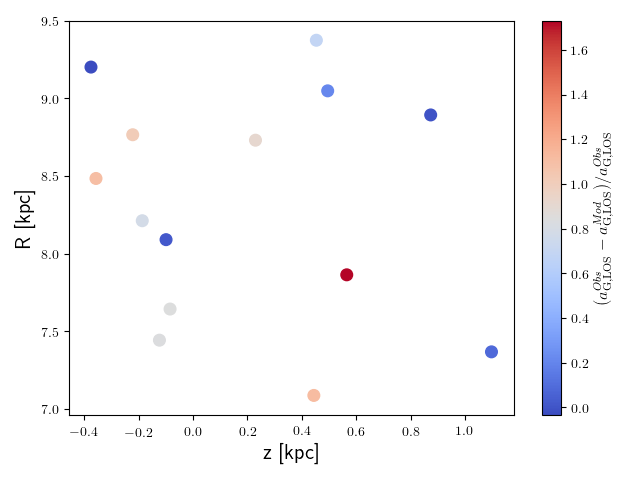}
\includegraphics[scale=0.5]{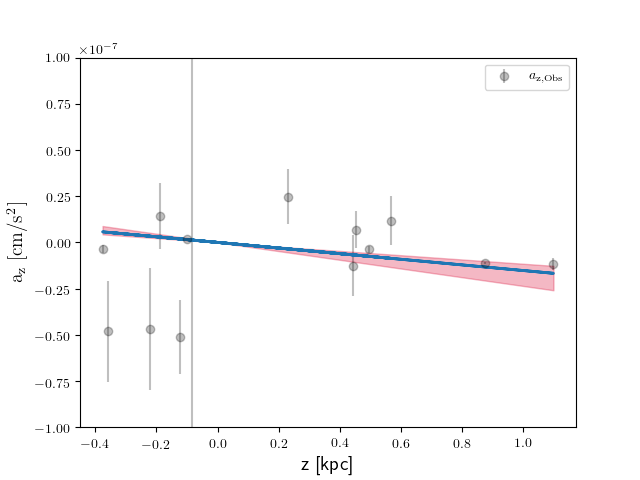}
\caption{(a)Residuals of the line-of-sight acceleration for the individual pulsars we analyze here relative to static models of the MW, as well as a local expansion and a polynomial fit. The different models are shown with different color points. (b) Residuals of the Galactic acceleration, $(a_{G,LOS}^{Obs} - a_{G, LOS}^{Mod})/a_{G,LOS}^{Obs}$, shown at the pulsar positions in R and z for the ($\alpha_{1},\beta$) model, with the colorbar displaying the values of the residuals. (c) The observed Galactic vertical acceleration compared to our fit for $a_{z}$, with the red shading showing the MCMC confidence intervals. \label{f:alos}}
\end{center}
\end{figure}

Figure \ref{f:alos} shows the fractional difference between the model line-of-sight acceleration $\amod$ for various static potentials and the observed values ($\aobs$) for all the pulsars in our sample.  Our focus will be on simple forms of the potential or low order expansions of the potential near the position of the Sun, as these pulsars cover a small area near the Sun. We express the potentials in terms of Galactocentric cylindrical radius $R=\sqrt{x^{2}+y^{2}}$ and $z$.
The static models that we  consider include a potential that is separable in the radial and vertical coordinates with potential $\Phi(R,z) = \Phi_R(R) + \Phi_z(z)$, as in \citet{Quillen2020}.   
The radial component may be written: 
\begin{equation}
\Phi_R (R) =  \left\{ \begin{array}{lll}
	V_{\rm LSR}^{2} \ln \left( \frac{R}{R_\odot} \right)   & {\rm for} & \beta=0 \\
	\frac{V_{\rm LSR}^{2}}{2 \beta} \left( \frac{R}{R_\odot} \right)^{2\beta}  & {\rm for} & \beta \ne 0. \\
\end{array}
\right. \label{eqn:Phi_R}
\end{equation}
where $V_{\rm LSR}$ is the local standard of rest rotational velocity $V_{\rm LSR} = 233.3 \pm 1.4~\rm km/s$ \citep{Schonrich2012}, and $R_{\odot}= 8.122 \pm 0.031~\rm kpc$ is the radial location of the Sun determined by the GRAVITY collaboration et al. (2018), and $\beta$ is the slope of the rotation curve, i.e., $\beta = \frac{d v_{c}}{dr} \rvert_{R_\odot}\frac{R_{\odot}}{V_{\rm  LSR}}$, where $v_{c}$ is the circular velocity.  We write the potential in the vertical direction as:
\begin{equation}
\Phi_{z}(z) = \frac{1}{2}\alpha_{1}z^{2} + \frac{1}{3}\alpha_{2}|z|^{3}
\label{eq:phiz}
\end{equation} 
and the components of the acceleration as: 
\begin{equation}
a_{R}=\frac{\partial}{\partial R} \Phi(R,z),~~~a_{z}=-\frac{\partial}{\partial z}\Phi(R,z)
\end{equation}
for an axisymmetric potential. For this and all other potentials, we fit for the vertical and radial accelerations simultaneously.  We refer to the $\beta=0$, $\alpha_2 = 0$ case as the $\alpha_{1}$ model, the $\beta=0$, $\alpha_2\neq 0$ as the ($\alpha_1,\alpha_2$) model, and the  $\beta\neq 0$, $\alpha_2=0$ case as the ($\alpha_{1}, \beta$) model in Table \ref{tab:Table1}. We also consider an exponential disk model of the form $\Phi = \rho_{0} \exp(-|z|/z_{0})$, as well as the Hernquist potential \citep{Hernquist1990}, where $M_{h}$ and $a_H$ are the mass normalization and scale length respectively for the Hernquist potential.  We also compare to the MWPotential2014 model that was presented by \citet{Bovy2015}, which is denoted ``MWP" in Table 1.  Finally, we consider a variant of the potential given in Eqs. \ref{eqn:Phi_R} and \ref{eq:phiz} and introduce a cross-term: 
\begin{equation}
\Phi(R,z) = V_{\rm LSR}^{2}~ \ln (R/R_{\odot}) + \ln(R/R_{\odot}) \gamma z^{2} + \frac{1}{2}\alpha_{1} z^{2}.
\end{equation}

This model assumes that the potential
is symmetric about the Galactic plane
and expands to second order in $z$.
To first order in $R - R_\odot$ we can
write $\ln (R/R_\odot) \sim (R-R_\odot)/R_\odot$.
We discuss below the sensitivity of $\gamma$ to the oblateness of the potential.  We refer to this model as the ``cross''-term model. 

We use the Markov-chain Monte Carlo (MCMC) code \textsc{emcee} (Foreman-Mackey et al. 2013) to explore the likelihood distribution of the data.  The log likelihood function is given by: 
\begin{align}
\log(L) = \log(P(\theta)) - \sum_i^N \frac{(\aobs - \amod)^{2}}{2\sigma_{i}^{2}}
\end{align}
where $\log(P(\theta))$ is the log prior on the parameters, $\theta$, $N$ is the number of pulsars, and $\sigma_{i}$ are the uncertainties.  The number of parameters used are $k+3N$, where $k$ is the number of parameters used in the various galactic models.  The three parameters that we use per pulsar are the parallax, e.g., distance, proper motion, $\mu$, and the secular GR effect, $\dot{P}_{\rm GR}$.  As these parameters have constraints on them, we use a log prior of the form $-(\theta_i - \theta_{i,\rm Obs})^2/\sigma_{i,\rm Obs}^2$, where $\sigma_{i,\rm Obs}$ is the published 1-$\sigma$ error on these measurements. For the $k$ parameters used in galactic models, we choose a flat distribution, but test its effects on our results.  Thus, in the MCMC calculation of the posterior distribution, we incorporate uncertainties in the measured $\PBDOTobs$ as well as uncertainties in terms that affect the calculation of the Shklovskii term (the distance and proper motion uncertainties) and the uncertainties in the calculation of $\PBDOTGR$ (i.e., the uncertainties on the mass of the pulsar and its companion and the eccentricity).  

\begin{figure}[ht]        
\begin{center}
\includegraphics[scale=0.55]{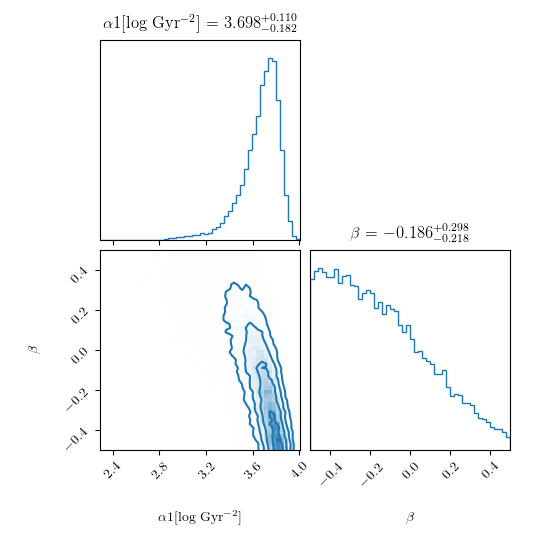}
\includegraphics[trim=0 0 0 0 , clip,scale=0.55]{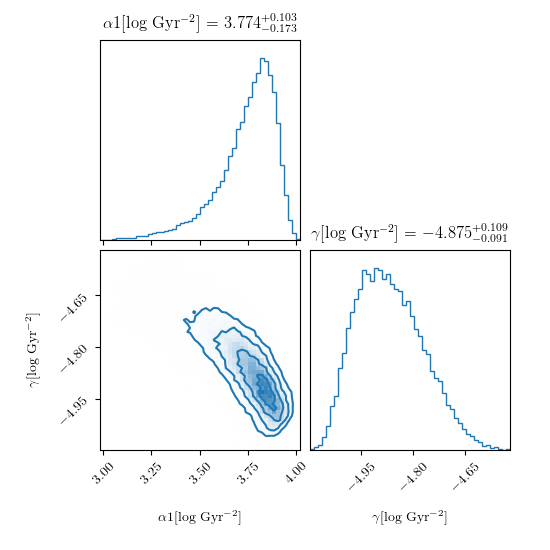}
\caption{
Top: Posterior probability distributions of $\alpha_{1}$ (which corresponds to the square of the frequency of low-amplitude vertical oscillations) and $\beta$ (the slope of the rotation curve). Bottom: posterior probability distribution of $\alpha_{1}$ and $\gamma$ (which is sensitive to oblateness).}
\label{f:corner}
\end{center}
\end{figure}

As shown in Figure \ref{f:alos}(a), the agreement between models and the observations are mostly within the errors of the measured uncertainties; those outside the measured uncertainties are within 2 $\sigma$. Table 2 gives the 16th, 50th, and 84th percentiles of the posterior probability distribution from the MCMC analysis, which reflects the uncertainty in the \PBDOTobs values, as well as the uncertainty in the parallaxes, proper motions, and in the masses and eccentricities.  Figure \ref{f:alos}(b) displays the residuals of the line-of-sight Galactic acceleration (having subtracted out the Shklovskii term and the GR term) at the pulsar positions in $R$ and $z$, for a representative model, the ($\alpha_{1},\beta$) model.  Figure \ref{f:alos}(b) shows that this model fits the data at the factor of $\sim$ 2 level in general, and that there are no clear patterns in the residuals.  A similar trend is observed for other models with comparable $\chi^{2}$ values.  Figure \ref{f:alos}(c) shows the observed vertical acceleration compared to the $(\alpha_{1},\beta)$ model.

To provide a measure of which models provide a better fit to the data, we also list here the Akaike Information Criteria (AIC; \cite{Ak1974}), which is given by:
\begin{equation}
\mathrm{AIC} = - 2\ln L + 2k    
\end{equation}
where $L$ is the likelihood and $k$ is the number of parameters in the model. The model with the lowest AIC is considered better at describing the data.  A $\Delta$AIC of 2 is considered positive evidence in favor of the model with the lower AIC, while a $\Delta$AIC of 6 indicates strong evidence \citep{KassRaferty1995}.  The $\alpha_1$, ($\alpha_1,\beta$), and ($\alpha_1,\gamma$) models all have a best fit value of $\log_{10}(\alpha_1/{\rm Gyr^{-2}})\approx 3.6-3.8$.  Our best-fit value for $\alpha_{1}$ (which describes the frequency of low-amplitude vertical oscillations) is close to a recent estimate by \cite{Quillen2020} to match the data presented from the Jeans analysis by \cite{HolmbergFlynn}.  

We may express a log-oblate (LO) potential with a core as:
\begin{equation}
\Phi_{\rm LO}(R,z) = \frac{V_{\rm LSR}^{2}}{2}\ln\left(\frac{R^{2}}{R_{\odot}^{2}} + \frac{z^{2}}{q^{2}R_{\odot}^{2}}+\frac{r_{\rm c}^{2}}{R_{\odot}^{2}}\right)   
\end{equation}
where $r_{\rm c}$ is the core size and $q < 1$ gives an oblate potential. A second-order expansion in $z$ and first-order expansion in $R$ about $R_\odot$ gives:
\begin{equation}
\gamma_{\rm LO}=-\frac{V_{\rm LSR}^{2}R_{\odot}^{2}}{\left(R_{\odot}^{2}+r_{c}^{2}\right)^{2}q^{2}}   
\end{equation}    
Evaluating this term for a log-spherical potential with $r_{\rm c}=0$ gives $\log_{10} (\gamma_{\rm LO}/{\rm Gyr}^{-2})=-2.93$, for $V_{\rm LSR}=233.3~\rm km/s$.  For the Miyamoto-Nagai (MN) disk:
\begin{equation}
\Phi_{\rm MN}(R,z) = \frac{-GM_{d}}{\sqrt{R^{2}+(a+\sqrt{z^{2}+b^{2}})^{2}}}
\end{equation}
where $M_d,a,b$ are the mass of the disk and the scale lengths respectively.  By expanding this potential to second order in $z$ near $z=0$ and to first order in $R$ near $R_{\odot}$, one can show that the  oblateness parameter for the Miyamoto-Nagai disk can be written as:
\begin{equation}
\gamma_{\rm MN}=-\frac{GM_{d}}{b}\frac{a+b}{(R_{\odot}^{2}+(a+b)^{2})^{5/2}}\frac{3R_{\odot}^{2}}{2}\end{equation}
Evaluating this quantity using the values listed in \citet{Candlish+2014}, i.e., $M_{d}=10^{11} M_{\odot}$, $b = 0.26$ kpc, $a = 6.5$ kpc, gives $ \log_{10}(\gamma_{\rm MN}/{\rm Gyr}^{-2}) = -3.94$, which is closer to our best-fit value for $\gamma$.  The oblateness inferred from pulsars is therefore consistent with that dominated by the disk potential and does not require a halo contribution, but which is consistent with expectations for a sample within a $\sim$ kpc of the Sun.

Figure \ref{f:corner} displays the posterior distribution for the ($\alpha_{1},\beta$) and ($\alpha_{1}, \gamma$) models.  We do not obtain constraints on $\beta$, the slope of the rotation curve, though the best-fit values are comparable to recent works \citep{Li+2019,Mroz2019}.  It is not surprising that we do not obtain a constraint for $\beta$ as our radial range is restricted to $\sim$ 1 kpc.  Expressed in dimensional terms, the slope is  $\approx -5^{6}_{-8}~\rm km/s/kpc$.

Models that are not symmetrical about the galactic plane (due to a warp or a lopsided mass distribution) or are non-axisymmetric may be constrained in future studies.  While our focus here has been in measuring the acceleration with a small sample of pulsars, direct acceleration measurements have the potential to provide a clear view of dark matter sub-structure for a sample of pulsars that are located at larger vertical heights, where the effects of interactions are more clearly manifest \citep{Chakrabarti2020}.

\subsection{The Oort limit from pulsar timing}
\label{sec:oort}

The Oort limit, or the volume mass density at the Galactic mid-plane, has traditionally been determined using kinematical tracers of the gravitational field \citep{Kuijken_Gilmore1989, HolmbergFlynn}, which assume spherical symmetry and equilibrium. Poisson's equation in cylindrical coordinates is $\frac{1}{R} \frac{\partial}{\partial R} R \frac{\partial}{\partial R} \Phi(R,z) + \frac{\partial^{2}}{\partial z^{2}} \Phi(R,z) = 4\pi G\rho_{0}$, which we evaluate at $z=0,R=R_{\odot}$. Using Eq. \ref{eqn:Phi_R} and \ref{eq:phiz} and Poisson's equation applied in the mid-plane at $R_{\odot}$, we can determine the frequency of low-amplitude vertical oscillations: 
\begin{align}
  \alpha_{1} = \nu^2 & =  4 \pi G \rho_0 - 2 \beta \Omega_\odot ^2
  \label{eqn:nu}
\end{align}
  where $\rho_0$ is the mid-plane mass density
  and we have used the potential of equation\;\ref{eqn:Phi_R} for the radial derivative terms.  In the special case of  $\beta=0$, $\alpha_{1} = 4 \pi G \rho_{0}$. Using the values of $\alpha_{1}$ and $\beta$ from Table 1, we obtain an Oort limit of 
  $0.08^{0.05}_{-0.02} M_{\odot}/\rm pc^{3}$.  This value of the Oort limit is close to, but somewhat lower relative to recent estimates using the Jeans equation \citep{McKee2015,HolmbergFlynn}.  Considering the baryon budget found by \cite{McKee2015} of $0.084 \pm 0.012~M_{\odot}/\rm pc^{3}$, we obtain a local dark matter density $\rho_{\rm DM} = -0.004^{0.05}_{-0.02}~M_{\odot}/\rm pc^{3}$, which is lower than, but within the range of prior work by \cite{McKee2015}, who found $\rho_{\rm DM}=0.013 \pm 0.003~\rm M_{\odot}/pc^{3}$.  It is close to but lower than the work by  \cite{Bovy_Tremaine2012}, who found $\rho_{\rm DM}=0.008 \pm 0.003 ~\rm M_{\odot}/pc^{3}$. It is also consistent with having no dark matter in the mid-plane.  Using the values of the baryon density from \cite{Bien2014} of $0.077 \pm 0.007~M_{\odot}/\rm pc^{3}$ gives $\rho_{\rm DM} = 0.0034^{0.05}_{-0.02}~M_{\odot}/\rm pc^{3}$. While the uncertainties on these values are large, our analysis does suggest that $\rho_{\rm DM}$ from the Jeans estimate may be an overestimate.  Improving the uncertainties on the Oort limit would allow us to directly determine the viability of disk dark matter models \citep{Randall2014}.  Recent work using \textit{Gaia} DR1 values by \cite{Schutz2018} using the Jeans analysis finds a local dark matter density of $0.038^{0.012}_{-0.015}~\rm M_{\odot}/\rm pc^{3}$ using A stars as tracers, $0.019^{0.012}_{-0.011}~\rm M_{\odot}/\rm pc^{3}$ using F stars as tracers, and $0.004^{0.01}_{-0.004}~\rm M_{\odot}/\rm pc^{3}$ using G stars as tracers.  Their value using G stars as tracers is consistent within the uncertainties to our value for the local dark matter density.

\section{Discussion}
\label{sec:discuss}

\begin{figure}[h]        
\begin{center}
\includegraphics[scale=0.5]{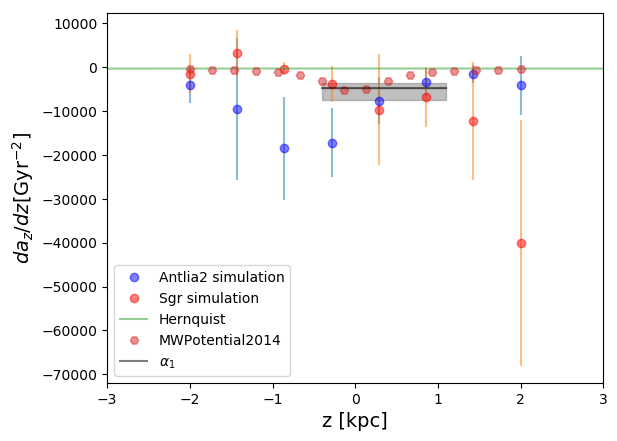}
\caption{The quantity $da_{z}/dz$ for the static potentials (Hernquist profile with $M=10^{12} M_{\odot}$ and $a_H = 30$ kpc, and MWPotential2014 \citep{Bovy2015}, and for the simulations of the Sgr dwarf and the Antlia 2 dwarf interacting with the Milky Way \citep{chakrabarti2019}, compared to our linear fit for $a_z$, which gives $da_{z}/dz = - \alpha_{1}$. The shaded regions display the current uncertainties on the fit.}
\label{f:sims}
\end{center}
\end{figure}

Figure \ref{f:sims} depicts a comparison of the quantity $da_{z}/dz$ from simulations of the Antlia 2 dwarf interacting with the Milky Way and the Sgr dwarf from \cite{chakrabarti2019}.  For our linear fitting function, $da_{z}/dz = -\alpha_{1}$. We compare to our value over the range of positions where we have analyzed pulsar timing data, along with the value for $da_{z}/dz$ for the static Hernquist potential with $M_{h} = 2 \times 10^{12} M_{\odot}$ and $a_H = 30~\rm kpc$, and for the MWPotential2014 model from \cite{Bovy2015}.  As the interacting simulations do not resolve the Solar neighborhood, we follow our earlier work \citep{Chakrabarti2020}, and calculate the acceleration in a ring of radius R = 8.1 kpc, as a function of $z$.  The average value is shown in the colored symbols, and the standard deviation along azimuth is shown in the error bars.  As is clear, $da_{z}/dz$ for interacting simulations varies in an asymmetric manner relative to the Galactic mid-plane, and as shown in our earlier work, this asymmetry develops as a result of the interaction with the dwarf galaxy.  The interactions that we consider here are due to fairly massive progenitor dwarf galaxies, with total masses $\sim 10^{10} M_{\odot}$ prior to the interaction.  A sample of pulsars at larger heights should be able to trace the asymmetry of $da_{z}/dz$, which is a signature of sub-structure, either due to interactions with dwarf galaxies, or dark matter sub-structure \citep{Chakrabarti2020}. 

We discuss here briefly additional implications of our work.  Pulsar timing measurements have been analyzed to constrain general relativity and alternate theories of gravity, most notably in the consistency of gravitational radiation \citep[e.g.,][]{Weisberg2016,Cameron+2018} but also in tests of the strong equivalence principle \citep[e.g.,][]{Freire+2012, Archibald+2018}, and the time-variability of the gravitational constant, $\dot{G}$ \citep{Damour+1988,Zhuetal2019}, while assuming a Galactic potential that is derived from kinematical analysis.  Obtaining high-precision RV measurements over ten year baselines towards lines-of-sight with pulsars can enable us to measure the $\Delta RV$ and thereby measure the Galactic acceleration via this complementary approach.  This measurement can then provide significantly more precise constraints on the parameters described above and constrain theories of gravity.  Although the uncertainties in fits for the time rate of change of the orbital period for binary pulsars due to gravitational radiation have improved (for the Hulse-Taylor system they are now within $\sim$ 1-sigma of the value predicted by relativity), they are currently dominated by the assumed values for the Galactic potential \citep{Weisberg2016}. Direct measurement of the potential would provide more robust constraints in these tests of gravity.

The solar acceleration has been measured by VLBI observations \citep{XuWang2012,Titov2013,Titov2018}.  \cite{Zakamska2005} have discussed the intriguing possibility of obtaining constraints on undiscovered planets or distant stellar companions from the acceleration of the solar system barycenter using pulsar timing observations.  The effect of a distant giant planet as in the work by \cite{Batygin2016} or that of the nearest stars is too small to affect our value of the Galactic acceleration, given current measurement uncertainties. 

\section{Conclusion}
\label{sec:conclusion}

We summarize our main findings below:

$\bullet$ By fitting a low-order polynomial for the Galactic potential to line-of-sight accelerations of fourteen binary pulsar systems, we infer an Oort limit of $0.08^{0.05}_{-0.02} M_{\odot}/\rm pc^{3}$.  Given the baryon budget from \cite{McKee2015}, this gives $\rho_{\rm DM} = -0.004^{0.05}_{-0.02}~M_{\odot}/\rm pc^{3}$; for the baryon budget from \cite{Bien2014}, $\rho_{\rm DM} = 0.0034^{0.05}_{-0.02}~M_{\odot}/\rm pc^{3}$. The Jeans analysis applied to \textit{Gaia} DR1 data gives $0.004^{0.01}_{-0.004}~\rm M_{\odot}/\rm pc^{3}$ for the local dark matter density, using G stars as tracers \citep{Schutz2018}.  The uncertainties in the local dark matter density are mainly due to the current uncertainty in the Oort limit from pulsar timing.  Higher precision measurements (not only for $\PBDOTobs$, but also for the distances and proper motions) would serve to reduce the uncertainty of this measurement.

$\bullet$ The vertical acceleration profile can be described by $a_{z} = -\alpha_{1} z$; our best-fit value for $\alpha_{1}$ is $\log_{10}( \alpha_{1}/{\rm Gyr}^{-2})=3.69^{0.19}_{-0.12}$. The posterior distribution of the slope of the rotation curve is not constrained, with $\beta = \frac{d v_{c}}{dr} \rvert_{R_\odot}\frac{R_{\odot}}{V_{\rm  LSR}} = -0.18^{+0.2}_{-0.3}$ (or expressed in dimensional terms is $\approx -5^{6}_{-8}~\rm km/s/kpc$).  The slope of the rotation curve could be measured in the future with a sample of pulsars at larger radial distances.

$\bullet$ The data imply an additional constraint on an oblateness parameter, $ \log_{10}( \gamma/{\rm Gyr}^{-2}) =-4.9^{0.1}_{-0.1}$.  This value of $\gamma$ is closer to that for disk models (which have larger $\gamma$) than halo models.  The oblateness inferred from pulsars is therefore consistent with that dominated by the disk potential and does not require a halo contribution, which is consistent with expectations for a sample within $\sim$ kpc of the Sun.

$\bullet$ Our analysis of dynamical simulations suggests that dark matter sub-structure or interactions with dwarf galaxies may manifest as asymmetries in $da_{z}/dz$ relative to a pure polynomial fit (such as our $\alpha_{1}$) or static models.  Nevertheless, the average value of $da_{z}/dz$ in the simulations we have considered here is close to our fit for $\alpha_{1}$.

$\bullet$ The measurement of the Galactic acceleration using high precision RV observations over ten year baselines near pulsars can provide significantly more precise constraints on $\dot{P}_b^{\rm GR}$, $\dot{G}$, and other post-Newtonian parameters than has been obtained thus far (for which prior work has assumed pre-formulated potentials that employ kinematic estimates).

\acknowledgments 
SC gratefully acknowledges support from the RCSA Time Domain Astrophysics Scialog award, NASA ATP NNX17AK90G, NSF AAG 2009574, and the IBM Einstein Fellowship at the Institute for Advanced Study.  PC is supported by the NASA ATP program through NASA grant NNH17ZDA001N-ATP. MTL and SJV are members of the NANOGrav project, which receives support from NSF Physics Frontiers Center award number 1430284. MTL also acknowledges support from NSF AAG 2009468.  SC thanks S. Tremaine, J. Goodman, and J. Wright for helpful discussions on the solar acceleration and R. Rafikov on pulsars, and C. McKee for helpful comments on the paper.  We also thank an anonymous referee for constructive feedback.

\bibliographystyle{aasjournal}
\bibliography{bibl}

\begin{thebibliography}{}
\expandafter\ifx\csname natexlab\endcsname\relax\def\natexlab#1{#1}\fi
\providecommand{\url}[1]{\href{#1}{#1}}
\providecommand{\dodoi}[1]{doi:~\href{http://doi.org/#1}{\nolinkurl{#1}}}
\providecommand{\doeprint}[1]{\href{http://ascl.net/#1}{\nolinkurl{http://ascl.net/#1}}}
\providecommand{\doarXiv}[1]{\href{https://arxiv.org/abs/#1}{\nolinkurl{https://arxiv.org/abs/#1}}}

\bibitem[{{Akaike}(1974)}]{Ak1974}
{Akaike}, H. 1974, IEEE Transactions on Automatic Control, 19, 716

\bibitem[{{Alam} {et~al.}(2020){Alam}, {Arzoumanian}, {Baker}, {Blumer},
  {Bohler}, {Brazier}, {Brook}, {Burke-Spolaor}, {Caballero}, {Camuccio},
  {Chamberlain}, {Chatterjee}, {Cordes}, {Cornish}, {Crawford}, {Cromartie},
  {DeCesar}, {Demorest}, {Dolch}, {Ellis}, {Ferdman}, {Ferrara}, {Fiore},
  {Fonseca}, {Garcia}, {Garver-Daniels}, {Gentile}, {Good}, {Gusdorff},
  {Halmrast}, {Hazboun}, {Islo}, {Jennings}, {Jessup}, {Jones}, {Kaiser},
  {Kaplan}, {Kelley}, {Shapiro Key}, {Lam}, {Lazio}, {Lorimer}, {Luo}, {Lynch},
  {Madison}, {Maraccini}, {McLaughlin}, {Mingarelli}, {Ng}, {Nguyen}, {Nice},
  {Pennucci}, {Pol}, {Ramette}, {Ransom}, {Ray}, {Shapiro-Albert}, {Siemens},
  {Simon}, {Spiewak}, {Stairs}, {Stinebring}, {Stovall}, {Swiggum}, {Taylor},
  {Tripepi}, {Vallisneri}, {Vigeland}, {Witt}, \& {Zhu}}]{Alam2020}
{Alam}, M.~F., {Arzoumanian}, Z., {Baker}, P.~T., {et~al.} 2020, arXiv
  e-prints, arXiv:2005.06495.
\newblock \doarXiv{2005.06495}

\bibitem[{{Archibald} {et~al.}(2018){Archibald}, {Gusinskaia}, {Hessels},
  {Deller}, {Kaplan}, {Lorimer}, {Lynch}, {Ransom}, \&
  {Stairs}}]{Archibald+2018}
{Archibald}, A.~M., {Gusinskaia}, N.~V., {Hessels}, J. W.~T., {et~al.} 2018,
  \nat, 559, 73, \dodoi{10.1038/s41586-018-0265-1}

\bibitem[{{Batygin} \& {Brown}(2016)}]{Batygin2016}
{Batygin}, K., \& {Brown}, M.~E. 2016, \aj, 151, 22,
  \dodoi{10.3847/0004-6256/151/2/22}

\bibitem[{{Bienaym{\'e}} {et~al.}(2014){Bienaym{\'e}}, {Famaey}, {Siebert},
  {Freeman}, {Gibson}, {Gilmore}, {Grebel}, {Bland-Hawthorn}, {Kordopatis},
  {Munari}, {Navarro}, {Parker}, {Reid}, {Seabroke}, {Siviero}, {Steinmetz},
  {Watson}, {Wyse}, \& {Zwitter}}]{Bien2014}
{Bienaym{\'e}}, O., {Famaey}, B., {Siebert}, A., {et~al.} 2014, \aap, 571, A92,
  \dodoi{10.1051/0004-6361/201424478}

\bibitem[{{Bovy}(2015)}]{Bovy2015}
{Bovy}, J. 2015, \apjs, 216, 29, \dodoi{10.1088/0067-0049/216/2/29}

\bibitem[{{Bovy}(2020)}]{Bovy2020}
---. 2020, arXiv e-prints, arXiv:2012.02169.
\newblock \doarXiv{2012.02169}

\bibitem[{{Bovy} \& {Tremaine}(2012)}]{Bovy_Tremaine2012}
{Bovy}, J., \& {Tremaine}, S. 2012, \apj, 756, 89,
  \dodoi{10.1088/0004-637X/756/1/89}

\bibitem[{{Burke-Spolaor} {et~al.}(2019){Burke-Spolaor}, {Taylor}, {Charisi},
  {Dolch}, {Hazboun}, {Holgado}, {Kelley}, {Lazio}, {Madison}, {McMann},
  {Mingarelli}, {Rasskazov}, {Siemens}, {Simon}, \&
  {Smith}}]{Burke-Spolaor+2019}
{Burke-Spolaor}, S., {Taylor}, S.~R., {Charisi}, M., {et~al.} 2019, \aapr, 27,
  5, \dodoi{10.1007/s00159-019-0115-7}

\bibitem[{{Cameron} {et~al.}(2018){Cameron}, {Champion}, {Kramer}, {Bailes},
  {Barr}, {Bassa}, {Bhandari}, {Bhat}, {Burgay}, {Burke-Spolaor}, {Eatough},
  {Flynn}, {Freire}, {Jameson}, {Johnston}, {Karuppusamy}, {Keith}, {Levin},
  {Lorimer}, {Lyne}, {McLaughlin}, {Ng}, {Petroff}, {Possenti}, {Ridolfi},
  {Stappers}, {van Straten}, {Tauris}, {Tiburzi}, \& {Wex}}]{Cameron+2018}
{Cameron}, A.~D., {Champion}, D.~J., {Kramer}, M., {et~al.} 2018, \mnras, 475,
  L57, \dodoi{10.1093/mnrasl/sly003}

\bibitem[{{Candlish} {et~al.}(2014){Candlish}, {Smith}, {Fellhauer}, {Gibson},
  {Kroupa}, \& {Assmann}}]{Candlish+2014}
{Candlish}, G.~N., {Smith}, R., {Fellhauer}, M., {et~al.} 2014, \mnras, 437,
  3702, \dodoi{10.1093/mnras/stt2166}

\bibitem[{{Chakrabarti} {et~al.}(2019){Chakrabarti}, {Chang}, {Price-Whelan},
  {Read}, {Blitz}, \& {Hernquist}}]{chakrabarti2019}
{Chakrabarti}, S., {Chang}, P., {Price-Whelan}, A.~M., {et~al.} 2019, \apj,
  886, 67, \dodoi{10.3847/1538-4357/ab4659}

\bibitem[{{Chakrabarti} {et~al.}(2020){Chakrabarti}, {Wright}, {Chang},
  {Quillen}, {Craig}, {Territo}, {D'Onghia}, {Johnston}, {De Rosa}, {Huber},
  {Rhode}, \& {Nielsen}}]{Chakrabarti2020}
{Chakrabarti}, S., {Wright}, J., {Chang}, P., {et~al.} 2020, arXiv e-prints,
  arXiv:2007.15097.
\newblock \doarXiv{2007.15097}

\bibitem[{{Chatterjee} {et~al.}(2009){Chatterjee}, {Brisken}, {Vlemmings},
  {Goss}, {Lazio}, {Cordes}, {Thorsett}, {Fomalont}, {Lyne}, \&
  {Kramer}}]{Chatterjeeetal2009}
{Chatterjee}, S., {Brisken}, W.~F., {Vlemmings}, W.~H.~T., {et~al.} 2009, \apj,
  698, 250, \dodoi{10.1088/0004-637X/698/1/250}

\bibitem[{{Cognard} {et~al.}(2017){Cognard}, {Freire}, {Guillemot}, {Theureau},
  {Tauris}, {Wex}, {Graikou}, {Kramer}, {Stappers}, {Lyne}, {Bassa},
  {Desvignes}, \& {Lazarus}}]{Cognard+2017}
{Cognard}, I., {Freire}, P. C.~C., {Guillemot}, L., {et~al.} 2017, \apj, 844,
  128, \dodoi{10.3847/1538-4357/aa7bee}

\bibitem[{{Damour} {et~al.}(1988){Damour}, {Gibbons}, \&
  {Taylor}}]{Damour+1988}
{Damour}, T., {Gibbons}, G.~W., \& {Taylor}, J.~H. 1988, \prl, 61, 1151,
  \dodoi{10.1103/PhysRevLett.61.1151}

\bibitem[{{Damour} \& {Taylor}(1991)}]{Damour_Taylor91}
{Damour}, T., \& {Taylor}, J.~H. 1991, \apj, 366, 501, \dodoi{10.1086/169585}

\bibitem[{{Deller} {et~al.}(2009){Deller}, {Bailes}, \& {Tingay}}]{Deller+2009}
{Deller}, A.~T., {Bailes}, M., \& {Tingay}, S.~J. 2009, Science, 323, 1327,
  \dodoi{10.1126/science.1167969}

\bibitem[{{Deller} {et~al.}(2013){Deller}, {Boyles}, {Lorimer}, {Kaspi},
  {McLaughlin}, {Ransom}, {Stairs}, \& {Stovall}}]{Deller+2013}
{Deller}, A.~T., {Boyles}, J., {Lorimer}, D.~R., {et~al.} 2013, \apj, 770, 145,
  \dodoi{10.1088/0004-637X/770/2/145}

\bibitem[{{Deller} {et~al.}(2019){Deller}, {Goss}, {Brisken}, {Chatterjee},
  {Cordes}, {Janssen}, {Kovalev}, {Lazio}, {Petrov}, {Stappers}, \&
  {Lyne}}]{Deller+2019}
{Deller}, A.~T., {Goss}, W.~M., {Brisken}, W.~F., {et~al.} 2019, \apj, 875,
  100, \dodoi{10.3847/1538-4357/ab11c7}

\bibitem[{{Desvignes} {et~al.}(2016){Desvignes}, {Caballero}, {Lentati},
  {Verbiest}, {Champion}, {Stappers}, {Janssen}, {Lazarus}, {Os{\l}owski},
  {Babak}, {Bassa}, {Brem}, {Burgay}, {Cognard}, {Gair}, {Graikou},
  {Guillemot}, {Hessels}, {Jessner}, {Jordan}, {Karuppusamy}, {Kramer},
  {Lassus}, {Lazaridis}, {Lee}, {Liu}, {Lyne}, {McKee}, {Mingarelli},
  {Perrodin}, {Petiteau}, {Possenti}, {Purver}, {Rosado}, {Sanidas}, {Sesana},
  {Shaifullah}, {Smits}, {Taylor}, {Theureau}, {Tiburzi}, {van Haasteren}, \&
  {Vecchio}}]{Desvignes+2016}
{Desvignes}, G., {Caballero}, R.~N., {Lentati}, L., {et~al.} 2016, \mnras, 458,
  3341, \dodoi{10.1093/mnras/stw483}

\bibitem[{{Fonseca} {et~al.}(2014){Fonseca}, {Stairs}, \&
  {Thorsett}}]{Fonseca+2014}
{Fonseca}, E., {Stairs}, I.~H., \& {Thorsett}, S.~E. 2014, \apj, 787, 82,
  \dodoi{10.1088/0004-637X/787/1/82}

\bibitem[{{Freire} {et~al.}(2012){Freire}, {Wex}, {Esposito-Far{\`e}se},
  {Verbiest}, {Bailes}, {Jacoby}, {Kramer}, {Stairs}, {Antoniadis}, \&
  {Janssen}}]{Freire+2012}
{Freire}, P. C.~C., {Wex}, N., {Esposito-Far{\`e}se}, G., {et~al.} 2012,
  \mnras, 423, 3328, \dodoi{10.1111/j.1365-2966.2012.21253.x}

\bibitem[{{Haines} {et~al.}(2019){Haines}, {D'Onghia}, {Famaey}, {Laporte}, \&
  {Hernquist}}]{Haines2019}
{Haines}, T., {D'Onghia}, E., {Famaey}, B., {Laporte}, C., \& {Hernquist}, L.
  2019, \apjl, 879, L15, \dodoi{10.3847/2041-8213/ab25f3}

\bibitem[{{Helmi} {et~al.}(2018){Helmi}, {Babusiaux}, {Koppelman}, {Massari},
  {Veljanoski}, \& {Brown}}]{helmi2018}
{Helmi}, A., {Babusiaux}, C., {Koppelman}, H.~H., {et~al.} 2018, \nat, 563, 85,
  \dodoi{10.1038/s41586-018-0625-x}

\bibitem[{{Hernquist}(1990)}]{Hernquist1990}
{Hernquist}, L. 1990, \apj, 356, 359, \dodoi{10.1086/168845}

\bibitem[{{Holmberg} \& {Flynn}(2000)}]{HolmbergFlynn}
{Holmberg}, J., \& {Flynn}, C. 2000, \mnras, 313, 209,
  \dodoi{10.1046/j.1365-8711.2000.02905.x}

\bibitem[{{Jennings} {et~al.}(2018){Jennings}, {Kaplan}, {Chatterjee},
  {Cordes}, \& {Deller}}]{Jenningsetal2018}
{Jennings}, R.~J., {Kaplan}, D.~L., {Chatterjee}, S., {Cordes}, J.~M., \&
  {Deller}, A.~T. 2018, \apj, 864, 26, \dodoi{10.3847/1538-4357/aad084}

\bibitem[{Kass \& Raftery(1995)}]{KassRaferty1995}
Kass, R.~E., \& Raftery, A.~E. 1995, Journal of the American Statistical
  Association, 90, 773.
\newblock \url{http://www.jstor.org/stable/2291091}

\bibitem[{{Kramer} {et~al.}(2006){Kramer}, {Stairs}, {Manchester},
  {McLaughlin}, {Lyne}, {Ferdman}, {Burgay}, {Lorimer}, {Possenti}, {D'Amico},
  {Sarkissian}, {Hobbs}, {Reynolds}, {Freire}, \& {Camilo}}]{Kramer+2006}
{Kramer}, M., {Stairs}, I.~H., {Manchester}, R.~N., {et~al.} 2006, Science,
  314, 97, \dodoi{10.1126/science.1132305}

\bibitem[{{Kuijken} \& {Gilmore}(1989)}]{Kuijken_Gilmore1989}
{Kuijken}, K., \& {Gilmore}, G. 1989, \mnras, 239, 605,
  \dodoi{10.1093/mnras/239.2.605}

\bibitem[{{Li} {et~al.}(2019){Li}, {Zhao}, \& {Yang}}]{Li+2019}
{Li}, C., {Zhao}, G., \& {Yang}, C. 2019, \apj, 872, 205,
  \dodoi{10.3847/1538-4357/ab0104}

\bibitem[{{Liu} {et~al.}(2020){Liu}, {Guillemot}, {Istrate}, {Shao}, {Tauris},
  {Wex}, {Antoniadis}, {Chalumeau}, {Cognard}, {Desvignes}, {Freire}, {Kehl},
  \& {Theureau}}]{Liu+2020}
{Liu}, K., {Guillemot}, L., {Istrate}, A.~G., {et~al.} 2020, arXiv e-prints,
  arXiv:2009.12544.
\newblock \doarXiv{2009.12544}

\bibitem[{{Lorimer} \& {Kramer}(2004)}]{Lorimer_Kramer2004}
{Lorimer}, D.~R., \& {Kramer}, M. 2004, {Handbook of Pulsar Astronomy}, Vol.~4

\bibitem[{{Manchester} {et~al.}(2005){Manchester}, {Hobbs}, {Teoh}, \&
  {Hobbs}}]{PSRCAT}
{Manchester}, R.~N., {Hobbs}, G.~B., {Teoh}, A., \& {Hobbs}, M. 2005, VizieR
  Online Data Catalog, VII/245

\bibitem[{{McKee} {et~al.}(2015){McKee}, {Parravano}, \&
  {Hollenbach}}]{McKee2015}
{McKee}, C.~F., {Parravano}, A., \& {Hollenbach}, D.~J. 2015, \apj, 814, 13,
  \dodoi{10.1088/0004-637X/814/1/13}

\bibitem[{{Mr{\'o}z} {et~al.}(2019){Mr{\'o}z}, {Udalski}, {Skowron}, {Skowron},
  {Soszy{\'n}ski}, {Pietrukowicz}, {Szyma{\'n}ski}, {Poleski}, {Koz{\l}owski},
  \& {Ulaczyk}}]{Mroz2019}
{Mr{\'o}z}, P., {Udalski}, A., {Skowron}, D.~M., {et~al.} 2019, \apjl, 870,
  L10, \dodoi{10.3847/2041-8213/aaf73f}

\bibitem[{{Phillips} {et~al.}(2020){Phillips}, {Ravi}, {Ebadi}, \&
  {Walsworth}}]{Phillips2020}
{Phillips}, D.~F., {Ravi}, A., {Ebadi}, R., \& {Walsworth}, R.~L. 2020, arXiv
  e-prints, arXiv:2008.13052.
\newblock \doarXiv{2008.13052}

\bibitem[{{Prager} {et~al.}(2017){Prager}, {Ransom}, {Freire}, {Hessels},
  {Stairs}, {Arras}, \& {Cadelano}}]{Prageretal2017}
{Prager}, B.~J., {Ransom}, S.~M., {Freire}, P. C.~C., {et~al.} 2017, \apj, 845,
  148, \dodoi{10.3847/1538-4357/aa7ed7}

\bibitem[{{Quillen} {et~al.}(2020){Quillen}, {Pettitt}, {Chakrabarti}, {Zhang},
  {Gagn{\'e}}, \& {Minchev}}]{Quillen2020}
{Quillen}, A.~C., {Pettitt}, A.~R., {Chakrabarti}, S., {et~al.} 2020, arXiv
  e-prints, arXiv:2006.01723.
\newblock \doarXiv{2006.01723}

\bibitem[{{Randall} \& {Reece}(2014)}]{Randall2014}
{Randall}, L., \& {Reece}, M. 2014, \prl, 112, 161301,
  \dodoi{10.1103/PhysRevLett.112.161301}

\bibitem[{{Ravi} {et~al.}(2019){Ravi}, {Langellier}, {Phillips}, {Buschmann},
  {Safdi}, \& {Walsworth}}]{Ravi2019}
{Ravi}, A., {Langellier}, N., {Phillips}, D.~F., {et~al.} 2019, \prl, 123,
  091101, \dodoi{10.1103/PhysRevLett.123.091101}

\bibitem[{{Read}(2014)}]{Read2014}
{Read}, J.~I. 2014, Journal of Physics G Nuclear Physics, 41, 063101,
  \dodoi{10.1088/0954-3899/41/6/063101}

\bibitem[{{Reardon} {et~al.}(2016){Reardon}, {Hobbs}, {Coles}, {Levin},
  {Keith}, {Bailes}, {Bhat}, {Burke-Spolaor}, {Dai}, {Kerr}, {Lasky},
  {Manchester}, {Os{\l}owski}, {Ravi}, {Shannon}, {van Straten}, {Toomey},
  {Wang}, {Wen}, {You}, \& {Zhu}}]{Reardon+2016}
{Reardon}, D.~J., {Hobbs}, G., {Coles}, W., {et~al.} 2016, \mnras, 455, 1751,
  \dodoi{10.1093/mnras/stv2395}

\bibitem[{{Reid} {et~al.}(2019){Reid}, {Menten}, {Brunthaler}, {Zheng}, {Dame},
  {Xu}, {Li}, {Sakai}, {Wu}, {Immer}, {Zhang}, {Sanna}, {Moscadelli}, {Rygl},
  {Bartkiewicz}, {Hu}, {Quiroga-Nu{\~n}ez}, \& {van Langevelde}}]{Reid2019}
{Reid}, M.~J., {Menten}, K.~M., {Brunthaler}, A., {et~al.} 2019, \apj, 885,
  131, \dodoi{10.3847/1538-4357/ab4a11}

\bibitem[{Sch{\"{o}}nrich(2012)}]{Schonrich2012}
Sch{\"{o}}nrich, R. 2012, Monthly Notices of the Royal Astronomical Society,
  427, 274, \dodoi{10.1111/j.1365-2966.2012.21631.x}

\bibitem[{{Sch{\"o}nrich} {et~al.}(2010){Sch{\"o}nrich}, {Binney}, \&
  {Dehnen}}]{Schonrich2010}
{Sch{\"o}nrich}, R., {Binney}, J., \& {Dehnen}, W. 2010, \mnras, 403, 1829,
  \dodoi{10.1111/j.1365-2966.2010.16253.x}

\bibitem[{{Schutz} {et~al.}(2018){Schutz}, {Lin}, {Safdi}, \&
  {Wu}}]{Schutz2018}
{Schutz}, K., {Lin}, T., {Safdi}, B.~R., \& {Wu}, C.-L. 2018, \prl, 121,
  081101, \dodoi{10.1103/PhysRevLett.121.081101}

\bibitem[{{Shapiro}(1964)}]{shapiro1964}
{Shapiro}, I.~I. 1964, Physical Review Letters, 13, 789,
  \dodoi{10.1103/PhysRevLett.13.789}

\bibitem[{{Shklovskii}(1970)}]{Shklovskii}
{Shklovskii}, I.~S. 1970, \sovast, 13, 562

\bibitem[{{Silverwood} \& {Easther}(2019)}]{Silverwood2019}
{Silverwood}, H., \& {Easther}, R. 2019, \pasa, 36, e038,
  \dodoi{10.1017/pasa.2019.25}

\bibitem[{{Titov} \& {Kr{\'a}sn{\'a}}(2018)}]{Titov2018}
{Titov}, O., \& {Kr{\'a}sn{\'a}}, H. 2018, \aap, 610, A36,
  \dodoi{10.1051/0004-6361/201731901}

\bibitem[{{Titov} \& {Lambert}(2013)}]{Titov2013}
{Titov}, O., \& {Lambert}, S. 2013, \aap, 559, A95,
  \dodoi{10.1051/0004-6361/201321806}

\bibitem[{{Weisberg} \& {Huang}(2016)}]{Weisberg2016}
{Weisberg}, J.~M., \& {Huang}, Y. 2016, \apj, 829, 55,
  \dodoi{10.3847/0004-637X/829/1/55}

\bibitem[{{Will}(2014)}]{Will2014}
{Will}, C.~M. 2014, Living Reviews in Relativity, 17, 4,
  \dodoi{10.12942/lrr-2014-4}

\bibitem[{{Xu} {et~al.}(2012){Xu}, {Wang}, \& {Zhao}}]{XuWang2012}
{Xu}, M.~H., {Wang}, G.~L., \& {Zhao}, M. 2012, \aap, 544, A135,
  \dodoi{10.1051/0004-6361/201219593}

\bibitem[{{Zakamska} \& {Tremaine}(2005)}]{Zakamska2005}
{Zakamska}, N.~L., \& {Tremaine}, S. 2005, \aj, 130, 1939,
  \dodoi{10.1086/444476}

\bibitem[{{Zhu} {et~al.}(2019){Zhu}, {Desvignes}, {Wex}, {Caballero},
  {Champion}, {Demorest}, {Ellis}, {Janssen}, {Kramer}, {Krieger}, {Lentati},
  {Nice}, {Ransom}, {Stairs}, {Stappers}, {Verbiest}, {Arzoumanian}, {Bassa},
  {Burgay}, {Cognard}, {Crowter}, {Dolch}, {Ferdman}, {Fonseca}, {Gonzalez},
  {Graikou}, {Guillemot}, {Hessels}, {Jessner}, {Jones}, {Jones}, {Jordan},
  {Karuppusamy}, {Lam}, {Lazaridis}, {Lazarus}, {Lee}, {Levin}, {Liu}, {Lyne},
  {McKee}, {McLaughlin}, {Os{\l}owski}, {Pennucci}, {Perrodin}, {Possenti},
  {Sanidas}, {Shaifullah}, {Smits}, {Stovall}, {Swiggum}, {Theureau}, \&
  {Tiburzi}}]{Zhuetal2019}
{Zhu}, W.~W., {Desvignes}, G., {Wex}, N., {et~al.} 2019, \mnras, 482, 3249,
  \dodoi{10.1093/mnras/sty2905}

\end{thebibliography}

\end{document}